\documentstyle[12pt,a4]{article}
\begin{document}
\hyphenation{anti-fermion}
\topmargin= -15mm
\textheight= 230mm
\baselineskip = 0.33 in
 \begin{center}
\begin{Large}

{\bf EDM OPERATOR FREE FROM 

SCHIFF'S THEOREM  }

\end{Large}

\vspace{1cm}

Tomoko ASAGA\footnote{JSPS fellow, Present address: 
Max-Planck-Institut f\"ur Kernphysik, 
Heidelberg, Germany},\footnote{e-mail: tomoko.asaga@mpi-hd.mpg.de} 
Takehisa FUJITA\footnote{e-mail: fffujita@phys.cst.nihon-u.ac.jp} and  
Makoto HIRAMOTO\footnote{e-mail: hiramoto@phys.ge.cst.nihon-u.ac.jp} 

Department of Physics, Faculty of Science and Technology  
  
Nihon University, Tokyo, Japan

\vspace{1cm}

{\large ABSTRACT} 

\end{center}

We present generalized Schiff's transformation on electric dipole moments (EDM) 
in quantum field theory. By the unitary transformation, 
 the time and parity violating 
interaction $ i{ge\over 2} \bar \psi \sigma_{\mu \nu} \gamma_5 \psi F^{\mu \nu} $ 
is transformed into a new form, but its nonrelativistic reduction 
has a unique form, which is free from Schiff's theorem. 
The relativistic corrections to the new EDM operator turn out to be 
a small increase to the EDM as given by 
$ b_2 (\alpha Z )^2 $ with $b_2 \simeq 2 $. Therefore, 
the calculation of the EDM with nonrelativistic Hartree-Fock 
wave functions presents the most conservative but reliable estimation for the 
enhancement factor of the EDM in atoms.

\vspace{1cm}
\noindent

\newpage
\begin{enumerate}
\item{\bf Introduction}

Symmetries in physics are fundamental for understanding nature. 
The highest symmetry in quantum field theory is the CPT (
charge conjugation, parity and time reversal) invariance. 
Any field theoretical models should be consistent with the CPT theorem. 
The next highest symmetry may be the time reversal invariance.  
Indeed, the T-invariance is also kept well in most of the field 
theoretical models. 

It is of fundamental interest to find the T-violating 
interactions in nature. Electric dipole moments (EDM) of particles, nuclei 
and atoms  in ground state reveal the violation of the T-invariance. 
Until now, the upper limit of the neutron EDM is around 
$-(3\pm5)\times 10^{-26} \  {\rm e}\cdot {\rm cm}$ [1]. 

There have been many experimental efforts to measure the EDM of the 
atomic systems. The best example is found for the EDM of $^{199}{\rm Hg}$ [2]. 
In this case, however, one has to be careful for extracting the 
EDM of the electrons or nucleons from the 
measurement of the atomic EDM since there is Schiff's 
theorem [3]. This theorem states that the EDM of the atom is canceled out 
due to the symmetry restoration mechanism as long as the constituents are 
interacting through the electromagnetic interactions with the nonrelativistic 
kinematics. 

Due to the presence of Schiff's theorem, people have long believed 
that the calculation of the enhancement factor in atomic systems 
should be carried out by employing the relativistic many body wave functions. 
However, it is not clear to what extent 
the Dirac Hartree-Fock method for the relativistic many body 
wave functions can be reliable. 
In fact, the Dirac Hartree-Fock method may well have some 
conceptual difficulties since the relativistic many body theory 
should be treated by field theories. 
The relativistic quantum mechanics is well defined 
only for the one body case. 

In this sense, it is of particular importance to find a way 
to estimate the EDM enhancement factor which does not depend on 
the ambiguity arising from atomic wave functions. 

In this paper, we present a generalization of Schiff's transformation 
which is extended to the quantum field theory of electrodynamics. 
We show that the T and P-violating interaction 
$ ige \bar \psi \sigma_{\mu \nu} \gamma_5 \psi F^{\mu \nu} $ is 
transformed into a new form which has a unique shape of the nonrelativistic 
reduction. For this operator, we can reliably calculate the EDM 
of the atomic system with nonrelativistic Hartree-Fock wave functions. 

In particular, the estimation of the EDM with Dirac wave functions 
in hydrogen-like atoms shows that  
we can write the EDM $d_Z$ of the atoms with $Z$ in units of electric 
EDM $d_e$ as 
$$d_Z = {b_0b_1\over{\epsilon_0}}  (Z\alpha )^2 \left(1
+b_2(Z\alpha )^2 \right) d_e \eqno{(1.1)} $$
where $b_0,b_1$ and $\epsilon_0$ are dimensionless constants 
which are determined by the nonrelativistic wave functions.  
$b_2$ is a positive number around $b_2 \simeq 2$. 
Eq.(1.1) indicates that the relativistic corrections are 
of the order of $ (\alpha Z )^2 $, and therefore they cannot be very large. 
Further, the relativistic effects must be smaller for the Hartree-Fock 
case than the hydrogen-like atoms since the Hartree-Fock effects always 
push the electron of the last orbit outward and 
thus the relativity becomes less. 

On the other hand, many body effects on the EDM are very large. This is 
mainly due to the fact that the EDM operator is, roughly speaking, 
proportional to ${1\over{r^2}}$, and thus the many body wave functions 
which are pushed outward due to electron electron repulsions 
reduce the EDM expectation values by an order of magnitude 
compared to the hydrogen-like wave function. Thus, the careful treatment 
of the many body effects is much more important than the relativistic 
corrections to the EDM.

In this paper, we calculate the EDM 
for the hydrogen, Li and Cs atoms as examples and 
estimate the EDM with the nonrelativistic Hartree-Fock wave functions. 
In particular, it is found that the enhancement 
factor of Cs atom in comparison with the electron EDM $d_e$ 
is 91.2, which should be compared with the previous 
estimations of 100-150 [4-10].

This paper is organized in the following way. In the next section, 
we explain briefly Schiff's theorem in the nonrelativistic 
quantum mechanics. Then, section 3 treats the field theoretical 
version of the Schiff transformation. In section 4, we evaluate 
the Foldy-Wouthuysen transformation of the relativistic EDM 
operator and obtain a unique form of the EDM operator which 
has no influence from Schiff's theorem. In section 5, we carry out  the 
numerical evaluations of the  EDM with the hydrogen-like wave functions. 
In particular, we compare the calculations with nonrelativistic and 
relativistic wave functions.  
In section 6, we present our calculations of the EDM with  
nonrelativistic Hartree-Fock wave functions. Section 7 summarizes 
what we have clarified in this paper. 

\vspace{3cm}

\item{\bf Schiff's theorem in quantum mechanics}

Here, we first review the shielding mechanism which was initially 
found by Schiff [3]. We consider the hamiltonian $H_0^{NR}$ 
of the atomic system  which is 
interacting through the Coulomb force as well as the external electric field 
${\bf E}^{ext}$,
$$ H_0^{NR} = \sum_i^Z \left( { {\bf p}_i^2\over{2m_i}} -
e A_0 ({\bf r}_i) + e{\bf r}_i\cdot {\bf E}^{ext} \right) \eqno{(2.1)} $$
where $A_0 ({\bf r}_i)$ is given as 
$$ A_0 ({\bf r}_i) = {Ze\over{r_i}}-
{1\over 2} \sum_{j\not=i} {e\over{|{\bf r}_i-{\bf r}_j|}}
 . \eqno{(2.2)} $$
To this system, we consider the T-and P-violating interaction $H_{edm}^{NR}$ 
$$ H_{edm}^{NR}= - \sum_i {\bf d}_i \cdot {\bf E}_i  \eqno{(2.3)}  $$
where ${\bf E}_i$ is defined as
$$ {\bf E}_i=-\nabla_i  A_0 ({\bf r}_i)+{\bf E}^{ext} \eqno{(2.4)} $$

${\bf d}_i$ represents the EDM of the $i$-th particle and can be written 
as 
$$ {\bf d}_i = ge \mbox{\boldmath $\sigma_i$} \eqno{(2.5)} $$
where 
$\mbox{\boldmath $\sigma_i$}$ represents the spin operator of the i-th 
particle. 
$g$ is a constant and represents the strength of the T and P-violating 
interaction, and it has the dimension of the length. 

Schiff's theorem states that the total hamiltonian 
$H^{NR}=H_0^{NR}+H_{edm}^{NR}$ 
is related to the $H_0^{NR}$ by the unitary transformation 
( Schiff transformation ) in the following way
$$ H^{NR}=H_0^{NR}+H_{edm}^{NR} 
=\exp \left( -i\sum_i{1\over{e}} {\bf p}_i \cdot {\bf d}_i \right)
 H_0^{NR} 
\exp \left( i\sum_i {1\over{e}}{\bf p}_i \cdot {\bf d}_i \right) 
\eqno{(2.6a)} $$
since we can easily evaluate the righthand side of eq.(2.6a)
$$ \exp \left( -i\sum_i{1\over{e}} {\bf p}_i \cdot {\bf d}_i \right) H_0^{NR} 
\exp \left( i\sum_i {1\over{e}}{\bf p}_i \cdot {\bf d}_i \right) 
= H_0^{NR}-\sum_i {\bf d}_i \cdot  {\bf E}_i  + O(g^2) . 
\eqno{(2.6b)} $$
Therefore, up to the order of $g$, eq.(2.6a) holds. 
By the unitary transformation, one obtains the same energy spectrum  
between $H^{NR}$ and $H_0^{NR}$. Therefore, this means that the effect of the 
$H_{edm}^{NR}$ is absorbed into the original hamiltonian and thus cannot be 
observed at all. This is Schiff's theorem. 

In his paper, Schiff argued that the shielding of the EDM may be 
violated by the relativistic effects or some other effects like 
the addition of the strong interactions. 

In what follows, we consider the Schiff transformation  for the 
relativistic case. 

\vspace{3cm}

\item{\bf Schiff transformation in QED}

\begin{enumerate}
\item{\bf Schiff transformation in lagrangian formulation}

Now, we want to generalize Schiff's theorem to the field theory. 
Here, we only consider the QED. The lagrangian density 
for fermions interacting with the gauge field can be written as  

$$  {\cal L}_0 =  \bar \psi ( i \gamma_{\mu} D^{\mu} - m_0 ) \psi 
  -{1\over 4} F_{\mu\nu} F^{\mu\nu}  \eqno{(3.1)}  $$
where $D_{\mu}$ and $F_{\mu \nu}$ denote the covariant derivative 
and the field tensor, respectively and are given as
$$ D_{\mu} = \partial_{\mu}+ieA_{\mu} \eqno{(3.2a)} $$
$$ F_{\mu \nu}= \partial_{\mu}A_{\nu}-\partial_{\nu}A_{\mu} . 
\eqno{(3.2b)}  $$  

In addition to the above lagrangian, we consider the T-and P-violating 
interaction which may presumably be induced by the supersymmetric model 
with some soft breaking interactions[11]. 
The T-and P-violating effective interaction can be given as 
$$ {\cal L}_{edm}= -i {ge\over 2} \bar \psi \sigma_{\mu \nu}
 \gamma_5 \psi F^{\mu \nu} \eqno{(3.3)} $$
where $\sigma_{\mu \nu}$ is defined as 
$$ \sigma_{\mu \nu} = {i\over 2} (\gamma_{\mu}\gamma_{\nu}- 
\gamma_{\nu}\gamma_{\mu} ) . \eqno{(3.4)} $$ 

The corresponding hamiltonian of eq.(3.3) which depends on the electric 
field can be written as 
$$ H_{edm}^{R1}= -ge\gamma_0 
\mbox{\boldmath $\Sigma$}\cdot {\bf E}  \eqno{(3.5a)} $$
where we define the relativistic spin operator $\mbox{\boldmath $\Sigma$}$ as
$ \mbox{\boldmath $\Sigma$} = \gamma^0 
\gamma^5 \mbox{\boldmath $\gamma$} .   $

Here, if we take a simple nonrelativistic limit of this operator, 
the hamiltonian becomes  
$$ H_{edm}^{R1} \simeq -ge \mbox{\boldmath $\sigma$}\cdot {\bf E} . 
\eqno{(3.5b)} $$
Since the ${\bf d} $ is written as $ {\bf d} = ge 
\mbox{\boldmath $\sigma$} $, this just corresponds to the $H_{edm}^{NR}$. 

Now, we consider the following unitary transformation
$$ \psi' = \exp \left( i{g} \gamma_5 
p_{\mu}\gamma^{\mu}  \right) 
  \psi .  \eqno{(3.6)} $$
Under this  transformation, the total lagrangian density 
${\cal L}={\cal L}_0 +{\cal L}_{edm} $ becomes 
up to the order of $g$
$$ {{\cal L}} = {\cal L}_0 
-2i{g} \bar \psi \gamma_5 (p_{\mu}p^{\mu} -eA_{\mu}p^{\mu}) \psi . 
 \eqno{(3.7)} $$
This shows that the unitary transformation of eq.(3.6) completely
 cancels out the ${\cal L}_{edm}$ term of eq.(3.3), but  
 a new term $2i{g} \bar \psi \gamma_5 p_{\mu}p^{\mu}  \psi $
  appears. 

Now, we want to make the corresponding  hamiltonian. 
In this case, we should make the 
conjugate momentum $\Pi_{A_0}$ for the $A_0$ and $\Pi_{\psi}$ 
for the $\psi$. It is clear that the terms proportional to the 
$p_0$ disappear from the hamiltonian. 
Therefore, the new EDM hamiltonian becomes 
$$ H_{edm}^{R2}=2i{g}\gamma_0\gamma_5 (p_{0}^2-{\bf p}^{2}) . 
  \eqno{(3.8)} $$
This is a new EDM hamiltonian which should be compared with 
the original one in eq.(3.5a). We should also note that the similar 
EDM hamiltonian is obtained in ref.[12]. In this case, the $p_0^2$ 
term is missing since they treat it in terms of the hamiltonian. 
Instead, we treat the problem in a  covariant fashion. However, 
it is easy to see that the $p_0^2$ term does not contribute 
to the mixing of the wave functions since it does not depend 
on the coordinates. In this respect, we reproduce the result 
obtained in ref.[12] with the lagrangian formulation 
with the full relativistic covariance. 

The important point is that the new EDM hamiltonian is free from 
Schiff's theorem. Therefore, this term indeed contributes to the 
generation of the EDM. 

Here, we should comment on the physical meaning of the unitary 
transformation of eq.(3.6). Under the transformation, we obtain 
a new lagrangian density which does not couple to the external 
electromagnetic field. This is just the statement which 
corresponds to Schiff's theorem. Namely, the EDM can only 
be induced from the second order effect of the new hamiltonian 
as will be treated later in eq.(5.2). 

\vspace{0.5cm}

\item{\bf Schiff transformation in hamiltonian formulation}

Next, we show here the same type of proof for the Dirac hamiltonian 
since the EDM leftover can be also obtained from the Dirac 
hamiltonian. This method is similar to the above proof 
of the field theory, but the transformation operator is different from 
the field theory treatment, but is the same as the nonrelativistic Schiff 
transformation. 

The Dirac hamiltonian for fermions interacting with the Coulomb force 
can be written as
$$ H_0^R = \sum_i^N \left( {\bf p}_i \cdot \mbox{\boldmath $\alpha_i$} + 
m_i\gamma_i^0-
e A_0 ({\bf r}_i) + e{\bf r}_i\cdot {\bf E}^{ext}  \right) .  \eqno{(3.9)} $$
In this case, the EDM hamiltonian which corresponds 
to ${\cal L}_{edm}$ can be written as
$$ H_{edm}^{R1}=-ge \sum_i^N \gamma_i^5 \mbox{\boldmath $\gamma_i$}
\cdot  {\bf E}_i .  \eqno{(3.10)} $$
Here, we can rewrite the Hamiltonian $H_{edm}^{R2} $ as 
$$ H_{edm}^{R1}=-ge \sum_i^N  \mbox{\boldmath $\Sigma_i$}\cdot {\bf E}_i 
-ge \sum_i^N (\gamma_i^0-1) \mbox{\boldmath $\Sigma_i$}
 \cdot {\bf E}_i  . \eqno{(3.11)} $$
In this case, one can easily check that the first term of eq.(3.11) 
is canceled out by the Schiff transformation in the same way as 
the nonrelativistic case of eq.(2.6a), and the following equation 
holds up to the order of $g$[13]. 
$$ H_0^{R}-ge \sum_i^N (\gamma_i^0-1) \mbox{\boldmath $\Sigma_i$} 
\cdot {\bf E}_i=\exp \left( i\sum_i{g} {\bf p}_i \cdot 
\mbox{\boldmath $\Sigma$}_i \right)
 (H_0^{R}+H_{edm}^{R1} ) 
\exp \left(- i\sum_i {g}{\bf p}_i \cdot 
\mbox{\boldmath $\Sigma$}_i \right) . 
\eqno{(3.12)} $$
Therefore, the leftover is just the second 
term of eq.(3.11), and we define $H_{edm}^{R3}$ as 
$$  H_{edm}^{R3}=-ge \sum_i^N (\gamma_i^0-1)
 \mbox{\boldmath $\Sigma_i$} \cdot {\bf E}_i  .  \eqno{(3.13)} $$
 
In the next section, we will prove that the two terms (eq.(3.8) and eq.(3.13)) 
have the identical shape of the nonrelativistic reduction under the 
Foldy-Wouthuysen transformation.  This nonrelativistic EDM operator 
is free from Schiff's theorem. 

\end{enumerate}

\vspace{3cm}

\item{\bf Nonrelativistic reduction of EDM operator}

When we want to evaluate operators which are expressed in 
the relativistic form, we should prepare wave functions which are 
obtained relativistically. In atomic systems with high Z, the lowest 
state is indeed quite relativistic. 

However, the system we are treating is not the hydrogen-like atom but 
real atoms. As an example, let us consider the Cs atom. 
In this case, there are many electrons around, and 
the state for the last electron in the ground state is 6s state which 
is far from relativistic. Therefore, we should rather obtain 
the EDM operator which is reduced to the nonrelativistic form. 
Since the EDM operators obtained in the last section are free from 
Schiff's theorem, we should make the reliable nonrelativistic 
reduction of the operator. 

Here, we employ the Foldy-Wouthuysen transformation which is 
a unitary transformation. It is here noted that the nonrelativistic 
reduction should be done at the last stage. Otherwise, one often makes 
mistakes since the Foldy-Wouthuysen transformation does not 
necessarily commute with other unitary transformation. 
Since the nonrelativistic reduction is an approximation, 
one has to do it after one has made other unitary transformations. 

In the case of the EDM operator, we have obtained the two different 
hamiltonians, eqs.(3.8) and (3.13). Here, we rewrite them again as the 
one body operators, 
$$ H_{edm}^{R2}=2i{g}\gamma_0\gamma_5 (p_{0}^2-{\bf p}^{2}) 
  \eqno{(3.8)} $$
$$  H_{edm}^{R3}=-ge (\gamma^0-1)
 \mbox{\boldmath $\Sigma$} \cdot {\bf E}  .  \eqno{(3.13)} $$
They look very much different from each other. 
Now, we want to make the nonrelativistic reduction of the 
two operators. 
 
By the Foldy-Wouthuysen transformation which is the most reliable method 
for the nonrelativistic reduction,  
 we write the reduced hamiltonian 
after some repeated operations of the transformation [14], 
$$ H_{FW} = \gamma_0 \left(m+{{\cal O}^2\over{2m}}-{{\cal O}^4
\over{8m^3}}\right) + 
{\cal E}  -{1\over{8m^2}}[{\cal O},[{\cal O},{\cal E} ]] 
\eqno{(4.1)}  $$
where $\cal O$ and $\cal E$ denote the odd and even operators 
in the gamma matrix space. 

In the case of the EDM hamiltonian of eq.(3.8), 
the  $\cal O$ and $\cal E$ can be written as 
$$ {\cal O}=  \mbox{\boldmath $\alpha$}\cdot {\bf p} 
-2i{g}\gamma_0\gamma_5 {\bf p}^2 \eqno{(4.2a)} $$
$$ {\cal E} = -eA_0 + e{\bf r}\cdot {\bf E}^{ext}  . \eqno{(4.2b)} $$
On the other hand, for the EDM hamiltonian of eq.(3.13), 
the  $\cal O$ and $\cal E$ become 
$$ {\cal O}=  \mbox{\boldmath $\alpha$}\cdot {\bf p}  \eqno{(4.3a)} $$
$$ {\cal E} = -eA_0 -ge (\gamma^0-1)
 \mbox{\boldmath $\Sigma$} \cdot {\bf E}
 + e{\bf r}\cdot {\bf E}^{ext}  . \eqno{(4.3b)} $$
After some straightforward calculation, 
we obtain the nonrelativistic hamiltonian
for both of the cases of eq.(3.8) and eq.(3.13),
$$ V_{edm} = {ge\over{2m^2}}  \left[ (\mbox{\boldmath $\sigma$}
\cdot {\bf E}) \nabla^2 -\rho (r)
(\nabla \cdot \mbox{\boldmath $\sigma$}) 
-2({\bf E} \cdot \nabla ) (\mbox{\boldmath $\sigma$} 
\cdot \nabla ) \right] \eqno{(4.4)}  $$
where $\rho (r)$ and ${\bf E}$ are defined as 
$$ \rho (r)=-\nabla^2 A_0(r) , \eqno{(4.5a)} $$
$$ {\bf E}= -\nabla A_0 (r)+{\bf E}^{ext}, \eqno{(4.5b)} $$ 
$$ A_0(r) = {Ze\over{r}} + \int {\rho_0 (r') \over{|{\bf r} 
-{\bf r}' | }} d^3 r'  \eqno{(4.5c)} $$
with $\rho_0 (r) = -e\sum_n |\psi_n (r)|^2 $.  

This is a proof that the apparent two different shapes 
of the relativistic hamiltonian eqs.(3.8) and (3.13) reduce 
to the identical and thus unique nonrelativistic hamiltonian.  

As we stressed before, this hamiltonian is free from Schiff's theorem. 
Therefore, there is definitely some mixture due to this operator 
between the opposite parity states like the 6s and 6p states.

\vspace{3cm}

\item{\bf EDM with hydrogen-like wave functions}

Since we obtain the EDM operator which is free from Schiff's theorem, 
we can now reliably calculate the EDM in actual cases. 
Before going to the calculations with the Hartree-Fock wave functions 
(in section 6), 
we first present our calculations with the hydrogen-like wave functions. 
The hydrogen-like wave functions are obviously far from reality in atoms 
since the interactions between electrons are very important, and 
thus the wave functions in the hydrogen-like atoms are 
quite different from the Hartree-Fock wave functions. But the use of 
the hydrogen-like wave functions helps us understand the basic 
structure of the EDM operator. 
This is quite important since we can evaluate all the matrix elements 
analytically. In addition, it turns out that the EDM 
eveluated by the hydrogen-like atom wave functions presents the upper 
limit of the enhancement factor even though the overestimation is 
quite significant. 

Here, we should note that the energy eigenvalues are replaced 
by the estimation using Hartree-Fock wave function 
since otherwise it is in fact  
meaningless to use the energy eigenvalues of the hydrogen-like atoms which are 
degenerate between $ns_{1\over 2}$ and $np_{1\over 2}$ states. 

Here, we consider the atomic systems in which one electron 
is found in the outer shell like Li or Cs cases. 

Now, we prepare the $ns_{1\over 2}$ and $np_{1\over 2}$ 
state wave functions which can be written as 
$$ \psi_{ns_{1\over 2}}({\bf r}) ={1\over{\sqrt{4\pi}}}
 R_{ns}(r)\xi_{1\over 2} , \eqno{(5.1a)} $$
$$ \psi_{np_{1\over 2}}({\bf r})= R_{np}(r) 
{\mbox{\boldmath $\sigma$} \cdot {\bf \hat r}\over{\sqrt{4\pi}}}
\xi_{1\over 2} \eqno{(5.1b)}$$
where $ R_{ns}(r)$ and $ R_{np}(r)$ represent the radial part of the 
atomic wave functions. $\xi_{1\over 2}$ denotes the spin part. 

In this case, we obtain the effective EDM $ d_{Z}$ for $ns_{1\over 2}$ 
state in atoms as
$$ d_{Z}=-2e \sum_{n'} {<ns_{1\over 2}|\sum_i z_i |n'p_{1\over 2}>
<n'p_{1\over 2}|V_{edm}|ns_{1\over 2}>
\over{E_{ns_{1\over 2}}-E_{n'p_{1\over 2}} } } . \eqno{(5.2)} $$
Here, we only consider the $n'=n$ case since this gives a dominant 
contribution to the EDM.  Further, we consider the atomic system when 
only one electron is found in the outer shell. 
Therefore, the summation 
over $i$ is just one single particle only.  
Also, we neglect the term which depends on the $\rho$ in eq.(4.4) 
since it vanishes for the hydrogen-like atoms. 

Now, we first evaluate the angular parts of
 the matrix elements $<ns_{1\over 2}|z|np_{1\over 2}>$ 
and $<np_{1\over 2}|V_{edm}|ns_{1\over 2}>$, and obtain  
$$  <\psi_{ns}|z |\psi_{np}> ={1\over{3} }
<R_{ns}|r |R_{np}> \eqno{(5.3a)}$$
$$<\psi_{np}|V_{edm}|\psi_{ns}> = {g\over{2m^2}}{Z\alpha}  
\int_0^\infty R_{np}(r) \left[ {1\over{r^2}}\left({2\over r}
{dR_{ns}(r)\over{dr}}-{d^2R_{ns}(r)\over{dr^2}} \right) \right] r^2dr . 
\eqno{(5.3b)} $$

Now, we want to make the matrix elements dimensionless and therefore 
we write them as
$$ <ns_{1\over 2}|z|np_{1\over 2}>= b_0 {a_0\over Z} \eqno{(5.4a)} $$
$$<np_{1\over 2}|V_{edm}|ns_{1\over 2}> = b_1 {g\over{2m^2}}(Z\alpha) 
\left( {Z\over{a_0}} \right)^4 \eqno{(5.4b)} $$ 
where $a_0$ denotes the Bohr radius in hydrogen atom and is written as 
$$ a_0={\hbar^2\over{me^2}} . \eqno{(5.5)} $$

Further, we write the energy difference 
$\Delta E=E_{np_{1\over 2}}-E_{ns_{1\over 2}}$ as
$$ \Delta E = \epsilon_0 m (Z\alpha )^2  . \eqno{(5.6)} $$ 
In this case, we can write the EDM as 
$$d_Z = {b_0b_1\over{\epsilon_0}} (Z\alpha )^2 d_e .  \eqno{(5.7)} $$

In what follows, we estimate the values of the $b_0$ and $b_1$ 
which depend on the states. Here, before doing so, we should make a comment 
concerning the first order contribution of the $V_{edm}$ term 
which couples directly to the external field of ${\bf E}^{ext}$. 
This contribution is a few orders of magnitude smaller than that 
of eq.(5.2). However, in the case of hydrogen ground state, 
the contribution amounts to 17 \% of eq.(5.2). Therefore, 
we take into account the first order contribution to the hydrogen case. 

\begin{enumerate}
\item{\bf Li (Z=3) : nonrelativistic case} 

In this case, the electron is in the $2s_{1\over 2}$ orbit. 
For the $b_0$ and $b_1$, we can easily evaluate them and find 
$$ b_0=-\sqrt{3} \eqno{(5.8a)} $$
$$ b_1=-{1\over{2\sqrt{3}}} .  \eqno{(5.8b)} $$
In this case, we can evaluate the EDM assuming the energy difference 
from the estimation using the Hartree-Fock wave functions
$$ \Delta E = 0.135{m\alpha^2\over 2} . \eqno{(5.9a)} $$ 
Thus, we obtain 
$$\epsilon_0=7.5\times 10^{-3} . \eqno{(5.9b)} $$ 
Therefore, we finally obtain for the EDM value of Li case
$$ d_{Li}=0.032 d_e . \eqno{(5.10)} $$
This value should be compared with the one obtained by 
the Hartree-Fock calculation. The EDM expectation value 
with the hydrogen-like wave function overestimates the EDM value 
by a factor of 5 as we will see later.

\item{\bf Li (Z=3) : relativistic case} 

Here, we evaluate the EDM matrix element using Dirac wave functions 
which are solved in a pure Coulomb potential $V(r)=-{Z\alpha\over r}$. 
In this case, the wave function is specified by the quantum number $n$ 
and $\kappa$. The ${ns_{1\over 2}}$ state ($\kappa =-1$) can be 
written as 
$$  \Psi_{ns_{1\over2 }} = \left( \matrix {
\ \ {1\over{\sqrt{4\pi}}}G_{n, \kappa}\xi_{1\over 2} & \  \cr
\ \ i {\mbox{\boldmath $\sigma$} \cdot {\bf \hat r}\over{\sqrt{4\pi}}}
F_{n, \kappa}\xi_{1\over 2}       & \  } \right) . \eqno{(5.11a)} $$
Also, the ${np_{1\over 2}}$ state ($\kappa=1$) can be 
written as 
$$  \Psi_{np_{1\over2 }} = \left( \matrix { \ \ 
{\mbox{\boldmath $\sigma$} \cdot {\bf \hat r}\over{\sqrt{4\pi}}}
{\tilde G}_{n, \kappa}\xi_{1\over 2} & \  \cr
\ \ i {1\over{\sqrt{4\pi}}}
{\tilde F}_{n, \kappa}\xi_{1\over 2}       & \  } \right) . \eqno{(5.11b)} $$
The analytic expressions of the radial wave functions 
$ G_{n, \kappa}$, $F_{n, \kappa}$, 
${\tilde G}_{n, \kappa}$ and ${\tilde F}_{n, \kappa}$ are given in Appendix. 

In this case, we can first evaluate the angular parts of 
the EDM matrix elements as, 
$$ <\Psi_{ns_{1\over2 }}|z|\Psi_{np_{1\over2 }}>
=-{1\over 3}\int_0^\infty (G_{n,-1}{\tilde G}_{n,1}+
F_{n,-1}{\tilde F}_{n,1} ) r^3dr \eqno{(5.12a)} $$
$$ <\Psi_{ns_{1\over2 }}|V_{edm}|\Psi_{np_{1\over2 }}>
=-{2g Z\alpha }\int_0^\infty F_{n,-1}{\tilde F}_{n,1} dr . \eqno{(5.12b)} $$

Now, for Li case,  the electron is in the $n=2$ and $\kappa=-1$ state.  
After some calculations, we obtain for $2s_{1\over 2}-2p_{1\over 2}$ case,  
$$ <\Psi_{2s_{1\over 2}}|z|\Psi_{2p_{1\over 2}}>=-\sqrt{3}
\left(1-{5\over{12}}(\alpha Z)^2 
\right)\left({a_0\over Z}\right) \eqno{(5.13a)} $$
$$ <\Psi_{2s_{1\over 2}}|V_{edm}|\Psi_{2p_{1\over 2}}>=-{g\over{2m^2}} 
(Z\alpha )\left({Z\over{a_0}} \right)^4 {1\over{2\sqrt{3} }} 
\left( 1+{13\over{6}}(\alpha Z)^2  \right)  . \eqno{(5.13b)} $$
Thus, we obtain the EDM with the relativistic wave functions as
$$ d_{Li}^R=0.032 \left( 1+1.75 (\alpha Z)^2 \right) d_e . \eqno{(5.14)} $$

Therefore, we can see that the relativistic correction to the EDM 
expectation value is of the order of $(\alpha Z)^2$, and thus 
it is quite small in this case. 

\vspace{1cm}
\item{\bf Cs (Z=55) : nonrelativistic case} 

In this case, the electron is in the $6s_{1\over 2}$ orbit. 
For the $b_0$ and $b_1$, we can easily evaluate them and find 
$$ b_0=-3\sqrt{35} \eqno{(5.15a)} $$
$$ b_1=-{\sqrt{35}\over{486}} . \eqno{(5.15b)} $$
In this case, we can evaluate the EDM assuming the energy difference 
from the Hartree-Fock estimation
$$ \epsilon_0 = 1.31\times 10^{-5} . \eqno{(5.16)} $$
Therefore, we finally obtain for the EDM value 
$$ d_{Cs}=2.66 \times 10^{3} d_e . \eqno{(5.17)} $$
This value should be compared with the one obtained by 
the Hartree-Fock calculation. The EDM expectation value 
with the hydrogen-like wave function overestimates the EDM value 
by a factor of 30 as we will see later.

\item{\bf Cs (Z=55) : relativistic case} 

In this case, the electron is in the $n=6$ and $\kappa =-1$ states.  
After some calculations, we obtain for $6s_{1\over 2}-6p_{1\over 2}$ case. 
$$ <\Psi_{6s_{1\over2 }}|z|\Psi_{6p_{1\over2 }}>
=-3\sqrt{35}\left(1-{13\over{84}}(\alpha Z)^2 
\right)\left({a_0\over Z}\right) \eqno{(5.18a)} $$
$$ <\Psi_{6s_{1\over2 }}|V_{edm}|\Psi_{6p_{1\over2 }}>=-{g\over{2m^2}} 
(Z\alpha )\left({Z\over{a_0}} \right)^4 {\sqrt{35}\over{486}} 
\left( 1+{257\over{126}}(\alpha Z)^2  \right) . \eqno{(5.18b)}  $$
Thus, we obtain the EDM with the relativistic wave functions as
$$ d_{Cs}^R=2.66\times 10^{3}
 \left( 1+1.88 (\alpha Z)^2 \right) d_e . \eqno{(5.19)} $$

Therefore, we can see that the relativistic correction to the EDM 
expectation value is again of the order of $(\alpha Z)^2$, and thus 
it is 30 \% correction to the EDM. But we should note that 
the relativistic effect must be smaller for the relativistic 
Hartree-Fock case, and thus the relativistic correction to the EDM 
can be at most 30 \%. 

Here, we note that the contribution due to 
the next order term $(\alpha Z)^4 $ is less than 10 \%, and thus we 
do not have to worry about the higher order terms. 

\item{\bf Hydrogen atom}

Finally, we want to present our calculation for the hydrogen atom. 
In this case, there is no stable $2s_{1\over 2}$ state, and therefore 
there is no enhancement for the EDM. However, the hydrogen atom 
is best studied in many respects, and thus 
there might be some chance that one can observe the EDM 
for this case with very high precision. 

Obviously, we have the wave functions analytically, and thus 
we can calculate the EDM in a closed form. 
Since the $2s_{1\over 2}$ state is not stable,  
we consider the ground state ($1s_{1\over 2}$) which can be mixed  
with the $2p_{1\over 2}$ state. 

For the $b_0$ and $b_1$, we can easily evaluate them and find 
$$b_0={128\sqrt{6}\over{729}} \eqno{(5.20a)} $$ 
$$b_1=-{16\over{9\sqrt{6}} } . \eqno{(5.20b)} $$
Since the energy difference is obtained as
$$ \Delta E = {3\over{ 8}}{m\alpha^2}  \eqno{(5.21a)} $$ 
we obtain
$$\epsilon_0={3\over 8} . \eqno{(5.21b)} $$ 
Finally we obtain the EDM for the hydrogen
$$ d_{H}= -5.32\times 10^{-5} d_e . \eqno{(5.22a)} $$
which should be compared with Sandars calculation [5],
$$ d_{H}= -2\alpha^2 d_e=-10.7\times 10^{-5} d_e . \eqno{(5.22b)} $$
The difference between them is due to the fact that Sandars took 
into account all of the $np_{1\over 2}$ states while we consider 
only the first excited state of $2p_{1\over 2}$. 

This is quite small and therefore it is not very clear whether 
one can choose the hydrogen atom for the EDM experiment.

\end{enumerate}

\vspace{3cm}

\item{\bf EDM from Hartree-Fock wave functions}

In the previous section, we have presented the EDM calculations 
with hydrogen-like wave functions. Now, we evaluate the EDM 
with the nonrelativistic Hartree-Fock wave functions [15]. 

Unfortunately, we cannot obtain the Hartree-Fock wave functions 
analytically, and thus we have to carry out all of the calculations 
numerically. However, since the basic expressions are given 
in terms of the hydrogen-like wave functions, we can follow the same 
notations.

\begin{enumerate}
\item{\bf Li (Z=3) : Hartree-Fock wave function} 

In this case, the electron is in the $2s_{1\over 2}$ orbit. 
For the $b_0$ and $b_1$, we can evaluate them and find 
$$ b_0=-4.17  \eqno{(5.23a)} $$
$$ b_1=-0.029 . \eqno{(5.23b)}$$
Now the energy difference between $2s_{1\over 2}$ 
and $2p_{1\over 2}$ states is calculated to be 
$$ \Delta E = 0.135{m\alpha^2\over 2} . \eqno{(5.24a)} $$ 
Therefore, we obtain 
$$ \epsilon_0 = 7.5\times 10^{-3} .  \eqno{(5.24b)} $$
Thus, we finally obtain 
$$ d_{Li}^{HF}=0.0077 d_e . \eqno{(5.25)} $$
This value should be compared with the EDM value calculated 
from the hydrogen-like wave function
$$ d_{Li}=0.032 d_e . $$
The absolute value of the EDM is smaller for the Hartree-Fock calculation 
than the hydrogen-like case by a factor of 5. 

This is quite easy to understand  since the wave function 
for the Hartree-Fock wave function should be pushed outward compared with 
the pure Coulomb case due to the electron-electron repulsion.

\item{\bf Cs (Z=55) : Hartree-Fock wave function} 

In this case, the electron is in the $6s_{1\over 2}$ orbit. 
For the $b_0$ and $b_1$, we can evaluate them and find 
$$ b_0=-116  \eqno{(5.26a)} $$
$$ b_1=-6.39\times 10^{-5} .\eqno{(5.26b)} $$
The energy difference between $6s_{1\over 2}$ 
and $6p_{1\over 2}$ states is calculated to be
$$ \Delta E = 0.025{m\alpha^2\over 2} .  \eqno{(5.27a)} $$ 
Thus, we obtain 
$$ \epsilon_0 = 1.31\times 10^{-5} . \eqno{(5.27b)} $$
Therefore, we finally obtain 
$$ d_{Cs}^{HF}=91.2 d_e . \eqno{(5.28)} $$

This is the enhancement factor of the EDM in Cs atom from the 
nonrelativistic Hartree-Fock calculation. This should be 
compared with the EDM calculated by the Dirac Hartree-Fock 
wave function. They predict the EDM in Cs atom around 
$(100-150) d_e$ depending on the calculations [4-10]. 
This is slightly larger than the one calculated here. 
This is indeed reasonable since the relativistic effects 
in EDM must be of the order of $(\alpha Z)^2$ increase 
compared to the nonrelativistic evaluations. 

In any case, it is now clear that the EDM in atoms is indeed 
enhanced compared with the electron EDM $d_e$ by a large 
enhancement factor for the atom with relatively high $Z$. 
Therefore, it would be better to use the atomic systems 
to observe the EDM.  

\end{enumerate}
\vspace{3cm}

\item{\bf Conclusions}

We have presented the nonrelativistic reduction of the EDM operator 
which is free from Schiff's theorem. This is important since we know 
definitely that the effect of the T-and P-violating interaction 
can be well observed in the atomic systems. Further, the enhancement 
factor is indeed quite large for the Cs atom. 
There are two reasons of the enhancement of the EDM in atoms. 
The first one is that the EDM operator is practically proportional 
to ${1\over{r^2}}$ and thus the EDM expectation value become large 
for the large $Z$. As the second reason, the EDM becomes enhanced due to  
a very small energy difference  between the ground state 
and the first excited state which has an opposite parity. 
Since the Hartree-Fock wave function is quite reliable in atomic systems, 
the estimated value of the EDM with the nonrelativistic reduction is indeed 
reliable. Here, we show that the relativistic correction to the EDM 
expectation value is of the order of $(Z\alpha)^2$ and therefore is 
not very large. Since the relativistic correction tends to increase 
the absolute value of the EDM, the nonrelativistic evaluation 
gives a conservative enhancement factor for the EDM in atoms. 

Here, we should make comments as to what is the difference 
between the present approach and the Schiff's calculation. 
The basic point is that the nonrelativistic reduction should be made 
at the last stage. For example, the simplest EDM operator is 
obtained when we make the nonrelativistic reduction at the lagrangian 
level. In this case, the EDM operator becomes eq.(3.5$b$) as discussed before. 
This is the worst case since there is no EDM operator left due to 
Schiff's theorem. 

The next level of the approximation which we have not presented 
in this paper is that we make the Foldy-Wouthuysen transformation before 
the Schiff transformation. Namely, we make the Foldy-Wouthuysen transformation 
for eq.(3.5$a$). In this case, we obtain the nonrelativistic EDM 
operator which is indeed different from eq.(4.4) after we 
remove the term described in eq.(2.3) by  the Schiff transformation 
of eq.(2.6$b$). 

The procedure we employ in this paper is that we make the Schiff transformation 
at the lagrangian level and then make the  Foldy-Wouthuysen transformation 
to the EDM operator. In this case, it is found that we obtain the unique 
nonrelativistic EDM operator which is free from Schiff's theorem. 

Therefore, as long as we employ the EDM operator of eq.(4.4), then we 
can estimate the EDM in atomic systems quite reliably with the Hartree-Fock 
wave functions. 

Finally, we make a comment as to how the present calculation is 
related to the relativistic Hartree-Fock calculations. 
If we want to treat very high $Z$ atoms, 
then we have to carry out the calculations 
relativistically. In this case, the direct 
evaluation of eq.(3.8) or (3.13) should be done if one 
obtains reliable wave functions with relativistic Hartree-Fock 
method [16]. However, if one really has to treat many body problems 
relativistically, then one has to carry out the calculations 
field theoretically. But this must be an extremely difficult task.

\end{enumerate}

\vspace{3cm}

We thank K. Asahi and K. Yazaki  
for helpful discussions and comments. 
This work is supported in part by 
Japan Society for the Promotion of Science.

\vspace{3cm}
\underline{\bf References}
\vspace{0.5cm}

1. K.F.Smith et al., Phys. Lett. {\bf B234}, 191 (1990)

2. S.K.Lamoreaux et al., Phys. Rev. Lett. {\bf 59}, 2275 (1987)

3. L.I.Schiff, Phys. Rev. {\bf 132}, 2194 (1963)

4. P.G.H.Sandars, Phys. Lett. {\bf 14}, 194 (1965)

5. P.G.H.Sandars, Phys. Lett. {\bf 22}, 290 (1966); 
J. Phys.{\bf B1}, 511 (1968)

6. A.C.Hartley, E.Lindroth and A-M.Martensson-Pendrill, 
J. Phys.{\bf B23}, 3417 (1990)

7. R.M.Sternheimer, Phys. Rev. {\bf 183}, 112 (1969)

8. V.K.Ignatovich, Zh. Eksp. Theor. Fiz. {\bf 56}, 2019 (1969)

9. W.R.Johnson, D. S.Guo, M. Idrees and J. Sapirstein, 
Phys. Rev. {\bf A32}, 2093 (1985)

10. A.-M.Martensson-Pendrill and P. \"Oster, Phys. Scripta {\bf 36}, 444 (1987)

11. Y. Kizukuri and N. Oshimo, Phys. Rev. {\bf D46}, 3025 (1992) 

12. W.R.Johnson, D. S.Guo, M. Idrees and J. Sapirstein, 
Phys. Rev. {\bf A34}, 1043 (1986)

13. I.B.Khriplovich, S.K.Lamoreaux, {\it CP Violation Without Strangeness}, 
Springer 1997

14. J.D.Bjorken, S.D.Drell, {\it Relativistic Quantum Mechanics}, 
McGraw-Hill, Inc. 1964

15. R.D. Cowman, Computer Code, RCN CODE for Hartree-Fock solutions in atoms

16. T.M.Byrnes, V.A.Dzuba, V.V. Flambaum and D.W. Murray, physics/9811044

\newpage

{\bf \underline{Appendix}} 


Here, we present the radial parts of the Dirac wave functions 
in  a pure Coulomb potential $V(r)=-{Z\alpha\over{r}}$. 

\begin{enumerate}
\item{{\bf $2s_{1\over 2}$ and $2p_{1\over 2}$ states } }

$$ G_{2,-1}(x)=\frac{m^{3/2}}{\Gamma(2\gamma+1)}
\sqrt{\frac{2\lambda^{5}(1+W)\Gamma(2\gamma+2)}{Z\alpha(Z\alpha+\lambda)}}
 x^{\gamma-1}e^{-x/2} \left[-1-\left(-1-\frac{Z\alpha}{\lambda}\right)
 F(-1,2\gamma+1,x) \right] $$

$$ F_{2,-1}(x)=\frac{-m^{3/2}}{\Gamma(2\gamma+1)}
\sqrt{\frac{2\lambda^{5}(1-W)\Gamma(2\gamma+2)}{Z\alpha(Z\alpha+\lambda)}}
 x^{\gamma-1}e^{-x/2} \left[1-\left(-1-\frac{Z\alpha}{\lambda}\right)
 F(-1,2\gamma+1,x) \right] $$
and 
$$ {\tilde G}_{2,1}(x)=\frac{m^{3/2}}{\Gamma(2\gamma+1)}
\sqrt{\frac{2\lambda^{5}(1+W)\Gamma(2\gamma+2)}{Z\alpha(Z\alpha-\lambda)}} 
x^{\gamma-1}e^{-x/2} \left[-1-\left(1-\frac{Z\alpha}{\lambda}\right)
F(-1,2\gamma+1,x) \right] $$

$$ {\tilde F}_{2,1}(x)=\frac{-m^{3/2}}{\Gamma(2\gamma+1)}
\sqrt{\frac{2\lambda^{5}(1-W)\Gamma(2\gamma+2)}{Z\alpha(Z\alpha-\lambda)}} 
x^{\gamma-1}e^{-x/2} \left[1-\left(1-\frac{Z\alpha}{\lambda}\right)
F(-1,2\gamma+1,x) \right] $$
where 
$$ \gamma=\sqrt{1-(Z\alpha)^2} $$

$$ \lambda=\sqrt{1-\left(\frac{E}{m}\right)^2} $$

$$ W=\frac{E}{m}=\frac{1}{\sqrt{1+(\frac{Z\alpha}{1+\gamma})^2}} $$

$$ x=2m\lambda r $$
Here, the function $F(a,b,c)$ denotes the confluent hypergeometric 
function.

\newpage

\item{\bf{ $6s_{1\over 2}$ and $6p_{1\over 2}$ states}}

$$ G_{6,-1}(x)=\frac{m^{3/2}}{\Gamma(2\gamma+1)}
\sqrt{\frac{2\lambda^{5}(1+W)\Gamma(2\gamma+6)}{5!Z\alpha(Z\alpha+\lambda)}}
x^{\gamma-1}e^{-x/2} $$

$$ \qquad\times\left[-5F(-4,2\gamma+1,x)-
\left(-1-\frac{Z\alpha}{\lambda}\right)F(-5,2\gamma+1,x) \right] $$

$$ F_{6,-1}(x)=\frac{-m^{3/2}}{\Gamma(2\gamma+1)}
\sqrt{\frac{2\lambda^{5}(1-W)\Gamma(2\gamma+6)}{5!Z\alpha(Z\alpha+\lambda)}}
x^{\gamma-1}e^{-x/2} $$

$$ \qquad\times\left[5F(-4,2\gamma+1,x)-
\left(-1-\frac{Z\alpha}{\lambda}\right)F(-5,2\gamma+1,x) \right] $$
and 
$$ {\tilde G}_{6,1}(x)=\frac{m^{3/2}}{\Gamma(2\gamma+1)}
\sqrt{\frac{2\lambda^{5}(1+W)\Gamma(2\gamma+6)}{5!Z\alpha(Z\alpha-\lambda)}}
x^{\gamma-1}e^{-x/2} $$

$$ \qquad\times\left[-5F(-4,2\gamma+1,x)-
\left(1-\frac{Z\alpha}{\lambda}\right)F(-5,2\gamma+1,x) \right] $$

$$ {\tilde F}_{6,1}(x)=\frac{-m^{3/2}}{\Gamma(2\gamma+1)}
\sqrt{\frac{2\lambda^{5}(1-W)\Gamma(2\gamma+6)}{5!Z\alpha(Z\alpha-\lambda)}}
x^{\gamma-1}e^{-x/2} $$

$$ \qquad\times\left[5F(-4,2\gamma+1,x)-
\left(1-\frac{Z\alpha}{\lambda}\right)F(-5,2\gamma+1,x) \right] $$
where 
$$ \gamma=\sqrt{1-(Z\alpha)^2} $$

$$ \lambda=\sqrt{1-\left(\frac{E}{m}\right)^2} $$

$$ W=\frac{E}{m}=\frac{1}{\sqrt{1+(\frac{Z\alpha}{5+\gamma})^2}} $$


\end{enumerate}

\end{document}